\documentclass[a4paper,11pt]{article}
\usepackage[left=3.17cm,right=3.17cm,top=2.54cm,
headheight=0.5cm,headsep=0.54cm,bottom=2.54cm,footskip=0.79cm
]{geometry}

\usepackage[all]{xy}
\usepackage{amsmath,amsfonts,amssymb,amsthm,epsfig,amscd,comment,latexsym,psfrag}
\usepackage{nicefrac,xspace,tikz}

\usetikzlibrary{arrows}
\usetikzlibrary{decorations.markings}
\def\Z{\mathbb Z}

\def\1{{\bf{1}}}

\usepackage[pdfstartview=FitH]{hyperref}

\def\Donebox{\begin{tikzpicture}
\draw (0,0)rectangle(0.25,0.25);
\draw (0,0.25) -- (0.175,0.35) [-];
\draw [shift = {+(0.25,0)}](0,0.25) -- (0.175,0.35) [-];
\draw [shift = {+(0.25,-0.250)}](0,0.25) -- (0.175,0.35) [-];
\draw (0.175,0.35) -- (0.425,0.35) [-];
\draw (0.425,0.35) -- (0.425,0.1) [-];
\end{tikzpicture}}
\def\Dtwoboxy{\begin{tikzpicture}
\draw (0,0)rectangle(0.5,0.25);
\draw (0.25,0) -- (0.25,0.25) [-];
\draw (0,0.25) -- (0.175,0.35) [-];
\draw [shift = {+(0.25,0)}](0,0.25) -- (0.175,0.35) [-];
\draw [shift = {+(0.5,0)}](0,0.25) -- (0.175,0.35) [-];
\draw [shift = {+(0.5,-0.250)}](0,0.25) -- (0.175,0.35) [-];
\draw (0.175,0.35) -- (0.675,0.35) [-];
\draw [shift = {+(0.25,0)}] (0.425,0.35) -- (0.425,0.1) [-];
\end{tikzpicture}}
\def\Dtwoboxz{\begin{tikzpicture}
\draw (0,0)rectangle(0.25,0.5);
\draw (0,0.25) -- (0.25,0.25) [-];
\draw [shift = {+(0,0.25)}](0,0.25) -- (0.175,0.35) [-];
\draw [shift = {+(0.25,0.25)}](0,0.25) -- (0.175,0.35) [-];
\draw [shift = {+(0.25,0)}](0,0.25) -- (0.175,0.35) [-];
\draw [shift = {+(0.25,-0.250)}](0,0.25) -- (0.175,0.35) [-];
\draw (0.425,0.625) -- (0.425,0.1) [-];
\draw [shift = {+(0.175,0.35)}] (0,0.25) -- (0.25,0.25) [-];
\end{tikzpicture}}
\def\Dtwoboxx{\begin{tikzpicture}
\draw (0,0)rectangle(0.25,0.25);

\draw (0,0.25) -- (0.35,0.45) [-];
\draw [shift = {+(0.25,0)}] (0,0.25) -- (0.35,0.45) [-];
\draw [shift = {+(0.25,-0.25)}] (0,0.25) -- (0.35,0.45) [-];
\draw [shift = {+(0.175,0.35)}](0,0) -- (0.25,0) [-];
\draw [shift = {+(0.25,-0.25)}][shift = {+(0.175,0.35)}](0,0) -- (0,0.25) [-];
\draw [shift = {+(0.35,0.45)}](0,0) -- (0.25,0) [-];
\draw [shift = {+(0.175,0.1)}] [shift = {+(0.25,-0.25)}][shift = {+(0.175,0.35)}](0,0) -- (0,0.25) [-];
\end{tikzpicture}}

\def\footnoterule{\kern 1mm \hrule width 7cm \kern 2.2mm}%

\newcommand{\bea}{\begin{eqnarray}}
\newcommand{\eea}{\end{eqnarray}}
\newcommand{\beaa}{\begin{eqnarray*}}
\newcommand{\eeaa}{\end{eqnarray*}}
\newcommand{\be}{\begin{equation}}
\newcommand{\ee}{\end{equation}}
\newcommand{\nn}{\nonumber}

\begin{document}

\title{3D Bosons and $W_{1+\infty}$ algebra}
\author{Wang Na\dag\footnote{Corresponding author: wangna@henu.edu.cn },\ Wu Ke\ddag\\
\dag\small School of Mathematics and Statistics, Henan University, Kaifeng, 475001, China\\
\ddag\small School of Mathematical Sciences, Capital Normal University, Beijing 100048, China}

\date{}
\maketitle

\begin{abstract}
In this paper, we consider 3D Young diagrams with at most $N$ layers in $z$-axis direction, which can be constructed by $N$ 2D Young diagrams on slice $z=j$, $j=1,2,\cdots, N$ from the Yang-Baxter equation. Use 2D Bosons $\{a_{j,m},\ m\in\Z\}$ associated to 2D Young diagrams on the slice $z=j$, we constructed 3D Bosons. Then we show the 3D Boson representation of $W_{1+\infty}$ algebra, and the Littlewood-Richardson rule for 3-Jack polynomials from the actions of 3D Bosons on 3D Young diagrams.
\end{abstract}

\section{Introduction}\label{sect1}
The Schur functions defined on 2D Young diagrams are an attractive research object, which were used to determine irreducible characters of highest weight representations of the classical groups, and the Littlewood-Richarson rule for Schur functions show the relations between the representation spaces\cite{FH,Mac,weyl}. There are many structures, such as 2D Bosons and Boson-Fermion correspondence, defined on Schur functions or 2D Young diagrams. These structures have many applications in mathematical physics. In \cite{MJD}, the group in the  Kyoto school uses Schur functions in a remarkable way to understand the KP and KdV hierarchies.  In \cite{NVT,PS}, Tsilevich and Su\l kowski, respectively, give the realization of the phase model in the algebra of Schur functions and build the relations between the $q$-boson model and Hall-Littlewood functions. In \cite{wang}, we build the relations between the statistical models, such as phase model, and KP hierarchy by using 2D Young diagrams and Schur functions. In \cite{WLZZ}, the authors show that the states in the $\beta$-deformed Hurwitz-Kontsevich matrix model can be represented as the Jack polynomials.

3D Young diagrams (plane partitions) are a generalization of 2D Young diagrams, which arose naturally in crystal melting model\cite{ORV,NT}. 3D Young diagrams also have many applications in many fields of mathematics and physics, such as statistical models, number theory, representations of some algebras (Ding-Iohara-Miki algebras, affine Yangian, etc). In this paper, we consider 3D Bosons and the Littlewood-Richardson rule for 3-Jack polynomials on 3D Young diagrams which parallel to 2D Bosons and the Littlewood-Richardson rule for Schur functions or Jack polynomials on 2D Young diagrams.

Let $a_{j,n}$ be the 2D Bosons associated to 2D Young diagrams which are on the slice $z=j$ of 3D Young diagrams with the relation
\[
[a_{j,n},a_{i,m}]=-\frac{1}{h_1h_2}\delta_{i,j}n\delta_{n+m,0},
\]
where $h_1,h_2$ are the parameters in the affine Yangian of $\mathfrak{gl}(1)$.
3D Bosons $b_{n,j}$ can be represented by these 2D Bosons. We treat 3D Young diagrams which have one layer in $z$-axis direction as 2D Young diagrams. Since we require 3D Bosons become 2D Bosons when $N=1$, which means 3D Bosons $b_{n,1}$ become 2D Bosons and $b_{n,j\geq 2}$ become zero, we know that 
$\langle 0|b_{n,j\geq 2}b_{-n,j\geq 2}|0\rangle$ must contain the factor $1+h_1h_2\psi_0$, where $\psi_0=-\frac{N}{h_1h_2}$ is the generator in affine Yangian of $\mathfrak{gl}(1)$. Since all the results we constructed on 3D Young diagrams are symmetric about three coordinate axes, which means they are symmetric about the parameters $h_1,h_2,h_3$, then $\langle 0|b_{n,j\geq 2}b_{-n,j\geq 2}|0\rangle$ must contain the factor
\[
(1+h_1h_2\psi_0)(1+h_1h_3\psi_0)(1+h_2h_3\psi_0).
\]

The Littlewood-Richardson rule for Schur functions are well known, for example,
\[
S_{(2)}S_{(1,1)}=S_{(3,1)}+S_{(2,1,1)}.
\]
They show the relations between the irreducible representation spaces of the general linear groups or permutation groups. In this paper, we calculate the Littlewood-Richardson rule for 3-Jack polynomials, we will find that it is more complicated than that for Schur functions, but it will become that for Schur functions in the special case $h_1=1,h_2=-1, N=1$. We believe that the Littlewood-Richardson rule should have applications in representation theory which we will consider later.

The paper is organized as follows. In section \ref{sect2}, we recall the definition of affine Yangian of $\mathfrak{gl}(1)$ and its representation on 3D Young diagrams. In section \ref{sect3},  we recall the definition of the $W_{1+\infty}$ algebra, then we construct the fields in $W_{1+\infty}$ algebra from the Miura transformation. The Virasoro field $V_2(z)$ become that in \cite{Pro1} when $h_1=h,h_2=-h^{-1}$. The spin $3$ field $V_3(z)$ is given new. In section \ref{sect4},  we construct 3D Boson fields and give the 3D Boson representation of $W_{1+\infty}$ algebra.
In section \ref{sect5}, we give the Littlewood-Richardson rule for 3-Jack polynomials. In section \ref{sect6}, we show the actions of 3D Bosons on 3D Young diagrams and the relations between 3D Bosons and the generators of affine Yangian of $\mathfrak{gl}(1)$.
\section{Affine Yangian of ${\mathfrak{gl}}(1)$}\label{sect2}
 In this section, we recall the definition of affine Yangian of ${\mathfrak{gl}}(1)$ and its representation on 3D Young diagrams. The affine Yangian $\mathcal{Y}$ of ${\mathfrak{gl}}(1)$ is an associative algebra with generators $e_j, f_j$ and $\psi_j$, $j = 0, 1, \cdots$ and the following relations\cite{Pro,Tsy}
\begin{eqnarray}
&&\left[ \psi_j, \psi_k \right] = 0,\\
&&\left[ e_{j+3}, e_k \right] - 3 \left[ e_{j+2}, e_{k+1} \right] + 3\left[ e_{j+1}, e_{k+2} \right] - \left[ e_j, e_{k+3} \right]\nonumber \\
&& \quad + \sigma_2 \left[ e_{j+1}, e_k \right] - \sigma_2 \left[ e_j, e_{k+1} \right] - \sigma_3 \left\{ e_j, e_k \right\} =0,\label{yangian1}\\
&&\left[ f_{j+3}, f_k \right] - 3 \left[ f_{j+2}, f_{k+1} \right] + 3\left[ f_{j+1}, f_{k+2} \right] - \left[ f_j, f_{k+3} \right] \nonumber\\
&& \quad + \sigma_2 \left[ f_{j+1}, f_k \right] - \sigma_2 \left[ f_j, f_{k+1} \right] + \sigma_3 \left\{ f_j, f_k \right\} =0, \label{yangian2}\\
&&\left[ e_j, f_k \right] = \psi_{j+k},\label{yangian3}\\
&& \left[ \psi_{j+3}, e_k \right] - 3 \left[ \psi_{j+2}, e_{k+1} \right] + 3\left[ \psi_{j+1}, e_{k+2} \right] - \left[ \psi_j, e_{k+3} \right]\nonumber \\
&& \quad + \sigma_2 \left[ \psi_{j+1}, e_k \right] - \sigma_2 \left[ \psi_j, e_{k+1} \right] - \sigma_3 \left\{ \psi_j, e_k \right\} =0,\label{yangian4}\\
&& \left[ \psi_{j+3}, f_k \right] - 3 \left[ \psi_{j+2}, f_{k+1} \right] + 3\left[ \psi_{j+1}, f_{k+2} \right] - \left[ \psi_j, f_{k+3} \right] \nonumber\\
&& \quad + \sigma_2 \left[ \psi_{j+1}, f_k \right] - \sigma_2 \left[ \psi_j, f_{k+1} \right] + \sigma_3 \left\{ \psi_j, f_k \right\} =0,\label{yangian5}
\end{eqnarray}
together with boundary conditions
\begin{eqnarray}
&&\left[ \psi_0, e_j \right]  = 0, \left[ \psi_1, e_j \right] = 0,  \left[ \psi_2, e_j \right]  = 2 e_j ,\label{yangian6}\\
&&\left[ \psi_0, f_j \right]  = 0,  \left[ \psi_1, f_j \right]  = 0,  \left[ \psi_2, f_j \right]  = -2f_j ,\label{yangian7}
\end{eqnarray}
and a generalization of Serre relations
\begin{eqnarray}
&&\mathrm{Sym}_{(j_1,j_2,j_3)} \left[ e_{j_1}, \left[ e_{j_2}, e_{j_3+1} \right] \right]  = 0, \label{yangian8} \\
&&\mathrm{Sym}_{(j_1,j_2,j_3)} \left[ f_{j_1}, \left[ f_{j_2}, f_{j_3+1} \right] \right]  = 0,\label{yangian9}
\end{eqnarray}
where $\mathrm{Sym}$ is the complete symmetrization over all indicated indices which include $6$ terms.

The notations $\sigma_2,\ \sigma_3$ in the definition of affine Yangian are functions of three complex numbers $h_1, h_2$ and $h_3$:
\begin{eqnarray}
\sigma_1 &=& h_1+h_2+h_3=0,\\
\sigma_2 &=& h_1 h_2 + h_1 h_3 + h_2 h_3,\\
 \sigma_3 &=& h_1 h_2 h_3.
\end{eqnarray}

The affine yangian $\mathcal{Y}$ has a representation on the plane partitions. A plane partition $\pi$ is a 2D Young diagram in the first quadrant of plane $xOy$ filled with non-negative integers that form nonincreasing rows and columns  \cite{FW, ORV}. The number in the position $(i,j)$ is denoted by $\pi_{i,j}$
\[
\left( \begin{array}{ccc}
\pi_{1,1} & \pi_{1,2} &\cdots \\
\pi_{2,1} & \pi_{2,2} &\cdots \\
\cdots & \cdots &\cdots \\
\end{array} \right).
\]
 The integers $\pi_{i,j}$ satisfy
\[
\pi_{i,j}\geq \pi_{i+1,j},\quad \pi_{i,j}\geq \pi_{i,j+1},\quad \lim_{i\rightarrow \infty}\pi_{i,j}=\lim_{j\rightarrow \infty}\pi_{i,j}=0
\]
for all integers $i,j\geq 0$. Piling $\pi_{i,j}$ cubes over position $(i, j)$ gives a 3D Young diagram. 3D Young diagrams arose naturally in the melting crystal model\cite{ORV,NT}. We always identify 3D Young diagrams with plane
partitions as explained above. For example, the 3D Young diagram
$\Dtwoboxy$
can also be denoted by the plane partition
$(1,1)$.

As in our paper \cite{3DFermionYangian}, we use the following notations. For a 3D Young diagram $\pi$, the notation $\Box\in \pi^+$ means that this box is not in $\pi$ and can be added to $\pi$. Here ``can be added'' means that when this box is added, it is still a 3D Young diagram. The notation $\Box\in \pi^-$ means that this box is in $\pi$ and can be removed from $\pi$. Here ``can be removed" means that when this box is removed, it is still a 3D Young diagram. For a box $\Box$, we let
\begin{equation}\label{epsilonbox}
h_\Box=h_1y_\Box+h_2x_\Box+h_3z_\Box,
\end{equation}
where $(x_\Box,y_\Box,z_\Box)$ is the coordinate of box $\Box$ in coordinate system $O-xyz$. Here we use the order $y_\Box,x_\Box,z_\Box$ to match that in paper \cite{Pro}.

Following \cite{Pro,Tsy}, we introduce the generating functions:
\begin{eqnarray}
e(u)&=&\sum_{j=0}^{\infty} \frac{e_j}{u^{j+1}},\nonumber\\
f(u)&=&\sum_{j=0}^{\infty} \frac{f_j}{u^{j+1}},\\
\psi(u)&=& 1 + \sigma_3 \sum_{j=0}^{\infty} \frac{\psi_j}{u^{j+1}},\nonumber
\end{eqnarray}
where $u$ is a parameter.
Introduce
\begin{equation}\label{psi0}
\psi_0(u)=\frac{u+\sigma_3\psi_0}{u}
\end{equation}
and
\begin{eqnarray} \label{dfnvarphi}
\varphi(u)=\frac{(u+h_1)(u+h_2)(u+h_3)}{(u-h_1)(u-h_2)(u-h_3)}.
\end{eqnarray}
For a 3D Young diagram $\pi$, define $\psi_\pi(u)$ by
\begin{eqnarray}\label{psipiu}
\psi_\pi(u)=\psi_0(u)\prod_{\Box\in\pi} \varphi(u-h_\Box).
\end{eqnarray}
In the following, we recall the representation of the affine Yangian on 3D Young diagrams as in paper \cite{Pro} by making a slight change. The representation of affine Yangian on 3D Young diagrams is given by
\begin{eqnarray}
\psi(u)|\pi\rangle&=&\psi_\pi(u)|\pi\rangle,\\
e(u)|\pi\rangle&=&\sum_{\Box\in \pi^+}\frac{E(\pi\rightarrow\pi+\Box)}{u-h_\Box}|\pi+\Box\rangle,\label{eupi}\\
f(u)|\pi\rangle&=&\sum_{\Box\in \pi^-}\frac{F(\pi\rightarrow\pi-\Box)}{u-h_\Box}|\pi-\Box\rangle\label{fupi}
\end{eqnarray}
where $|\pi\rangle$ means the state characterized by the 3D Young diagram $\pi$ and the coefficients
\begin{equation}\label{efpi}
E(\pi \rightarrow \pi+\square)=-F(\pi+\square \rightarrow \pi)=\sqrt{\frac{1}{\sigma_3} \operatorname{res}_{u \rightarrow h_{\square}} \psi_\pi(u)}
\end{equation}
 Equations (\ref{eupi}) and (\ref{fupi}) mean generators $e_j,\ f_j$ acting on the 3D Young diagram $\pi$ by
\begin{equation}
\begin{aligned}
e_j|\pi\rangle &=\sum_{\square \in \pi^{+}} h_{\square}^j E(\pi \rightarrow \pi+\square)|\pi+\square\rangle,
\end{aligned}
\end{equation}
\begin{equation}
\begin{aligned}
f_j|\pi\rangle &=\sum h_{\square}^j F(\pi \rightarrow \pi-\square)|\pi-\square\rangle .
\end{aligned}
\end{equation}

In the following of this paper, we consider 3D Young diagrams which have at most $N$ layers in the $z$-axis direction, and slice the 3D Young diagrams into a series of 2D Young diagrams by the plane $z=n$ for $n=1,2,\cdots,N$. Then the symmetry of the affine Yangian of $\mathfrak{gl}(1)$ about the coordinate axes are broken. For example, $\psi_0=-\frac{N}{h_1h_2}$.
\section{$W_{1+\infty}$ algebra}\label{sect3}
We begin this section by the definition of $W_{1+\infty}$ algebra. The $W_{1+\infty}$ algebra contains the Heisenberg algebra, the Virasoro algebra as  subalgebras\cite{Pro}. The generators are $V_{j,m}$ for $j\in\Z_+$, $m\in\Z$. The relations are
\bea
[V_{1,m}, V_{1,n}]&=&m\delta_{m+n,0}c_{1},\eea
\bea
[V_{2,m}, V_{2,n}]&=&(m-n) V_{2,m+n}+\frac{m^3-m}{12} \delta_{m+n,0} c_{2},
\eea
\bea
[V_{2,m}, V_{1,n}]&=&-n V_{1,m+n},
\eea
and generally
\be
[V_{j,m}, V_{k,n}]=\sum\limits_{\substack{0\leq l\leq j+k-2 \\ j+k-l\text{even}}}C_{jk}^lN_{jk}^l(m,n)V_{l,m+n},
\ee
where the coefficients $N_{jk}^l(m,n)$ are
\bea
N_{jk}^0(m,n)&=&\left(\begin{array}{cc}m+j-1\\j+k-1\end{array}\right)\delta_{m+n,0},\\
N_{jk}^l(m,n)&=&\sum_{s=0}^{j+k-l-1}\frac{(-1)^s}{(j+k-l-1)!(2l)_{j+k-l-1}}\left(\begin{array}{cc}j+k-l-1\\s\end{array}\right)\times\nn\\
&&[j+m-1]_{j+k-l-1-s}[j-m-1]_{s}[k+n-1]_{s}[k-n-1]_{j+k-l-1-s},\nn
\eea
and the structure constants $C_{jk}^l$ are
\bea
C_{jk}^0&=&\frac{(j-1)!^2(2j-1)!}{4^{j-1}(2j-1)!!(2j-3)!!}\delta_{jk}c_{j},\\
C_{jk}^l&=&\frac{1}{2\times 4^{j+k-l-2}}(2l)_{j+k-l-1}\times {}_4F_3\left(\begin{array}{cc}\frac{1}{2},\frac{1}{2},-\frac{1}{2}(j+k-l-2),-\frac{1}{2}(j+k-l-1)\\ \frac{3}{2}-j,\frac{3}{2}-k,\frac{1}{2}+l\end{array};1\right),\nn
\eea
with
\bea
(a)_n&=&a(a+1)\cdots (a+n-1),\\
{[a]}_n&=&a(a-1)\cdots (a-n+1),\\
{}_mF_n\left(\begin{array}{cc}a_1,\cdots,a_m\\ b_1,\cdots,b_n\end{array};z\right)&=&\sum_{k=0}^\infty\frac{(a_1)_k\cdots (a_m)_k}{(b_1)_k\cdots (b_n)_k}\frac{z^k}{k!}.
\eea
Note that here we allow that the central charges can be different.

We consider $J_j(z)$
\be
J_j(z)=\sum_{n\in\Z}a_{j,n}z^{-n-1}
\ee
with the relation
\be\label{ajnakmcom}
[a_{j,n},a_{k,m}]=-\frac{1}{h_1h_2}\delta_{j,k}n\delta_{n+m,0}.
\ee
Define
\be\label{j(z)}
J(z)=\sum_{n\in\Z}a_{n}z^{-n-1}=J_1(z)+J_2(z)+\cdots+J_3(z),
\ee
then the bosons $a_{n}$ satisfy
\be
[a_{n},a_{m}]=-\frac{N}{h_1h_2}n\delta_{n+m,0}=\psi_0n\delta_{n+m,0}.
\ee
In the following two subsections, we consider the Boson $a_{j,n}$ representation of $W_{1+\infty}$ algebra.
\subsection{Miura transformation and the $W_{1+\infty}$ algebra $V_n(z)$}
Let
\[
\alpha_0=-\frac{h_3}{h_1h_2},
\]
and define the operator $U_k(z)$ as in \cite{Pro1} by
\be
:(\alpha_0\partial+J_1(z))(\alpha_0\partial+J_2(z))\cdots (\alpha_0\partial+J_N(z)):=\sum_{k=0}^N U_k(z)(\alpha_0\partial)^{N-k}.
\ee
The fields $U_k(z)$ generate an algebra, which is $W_{1+N}$. The fields $V_n(z)$ can be realized by $U_k(z)$. We list the concrete expressions of the first few $U_k(z)$ as in \cite{pro1411}
\begin{eqnarray}
U_0 & = & 1, \\
U_1 & = & \sum_{j=1}^N J_j, \\
U_2 & = & \sum_{j<k} :J_j J_k : + \alpha_0 \sum_{j=1}^N (j-1) J_j^\prime, \\
U_3 & = & \sum_{j<k<l} :J_j J_k J_l: + \alpha_0 \sum_{j<k} (j-1) :J_j^\prime J_k: \\
\nonumber
& & + \alpha_0 \sum_{j<k} (k-2) :J_j J_k^\prime: + \frac{\alpha_0^2}{2} \sum_{j=1}^N (j-1)(j-2) J_j^{\prime\prime}.
\end{eqnarray}
Note that the expressions of $U_k(z)$ is the same with that in \cite{pro1411}, but the commutation relation of Boson fields $J_j(z)$ are different from that in \cite{pro1411}.

Clearly, the Boson field
\bea
U_1(z)=J(z)=J_1(z)+J_2(z)+\cdots+J_N(z),\eea
is the same with that in \cite{pro1411} with a slight different commutation relation (or OPE)
\be
U_1(z)U_{1}(w)\sim -\frac{N}{h_1h_2}\frac{1}{(z-w)^2},
\ee
which means the central charge of Boson field $U_{1}(z)$ is $\psi_0$. Define $V_1(z)=U_1(z)$.

To get the Virasora field, we need the following OPE
\beaa
U_2(z) U_1(w) & \sim &-h_1h_2\left( \frac{-N(N-1)\alpha_0}{(z-w)^3} + \frac{(N-1)U_1(w)}{(z-w)^2} + \frac{(N-1)U_1^\prime(w)}{z-w}\right),\\
U_1^\prime (z)U_1(w)&\sim & -\frac{N}{h_1h_2}\frac{-2}{(z-w)^3},\\
U_1U_1(z)U_1(w)&\sim&-\frac{2N}{h_1h_2}\left(\frac{U_1(w)}{(z-w)^2}+\frac{U_1^\prime(w)}{(z-w)}\right),
\eeaa
where we use $AB(z)$ to denote the normal order $:A(z)B(z):$.
Note that these relations become that in \cite{pro1411} when $h_1=h, h_2=-h^{-1}$.
Then
\be\label{v2z}
V_2(z)=-h_1h_2\left( -U_2(z) + \frac{(N-1)\alpha_0}{2} U_1^\prime(z) + \frac{1}{2} (U_1 U_1)(z)\right),
\ee
which satisfy
\be
V_2(z)U_1(w)\sim \frac{U_1(w)}{(z-w)^2}+\frac{U_1^\prime(w)}{(z-w)}.
\ee
We also have
\bea
V_2(z)&=&-\frac{h_1h_2}{2}\sum_{j=1}^N :J_j(z)J_j(z):+\frac{h_3}{2}\sum_{j=1}^N(N+1-2j) J_j^\prime(z).\label{v2zj}
\eea
which equals the special case $h_\epsilon=h_3$ in \cite{WBCW}, and the following OPE
\be
V_2(z)V_2(w)\sim \frac{c_2/2}{(z-w)^4}+\frac{2V_2{(w)}}{(z-w)^2}+\frac{V_2^\prime(w)}{z-w},
\ee
with
\be
c_2=N+h_1h_2\alpha_0^2N(N+1)(N-1).
\ee
As in paper \cite{Pro1}, let
\[
\lambda_1=-\psi_0 h_2h_3, \ \lambda_2=-\psi_0h_1h_3,\ \lambda_3=-\psi_0 h_1h_2,
\]
the central charge $c_2$ equals
\bea
c_2&=&1+(\lambda_1-1)(\lambda_2-1)(\lambda_3-1)\nn\\
&=&1-(1+\psi_0h_1h_2)(1+\psi_0h_1h_3)(1+\psi_0h_2h_3)\nn\\
&=&-(\psi_0\sigma_2+\psi_0^3\sigma_3^2).
\eea
Note that the central charge $c_2$ here is the same with that in \cite{pro1411,Pro1}, but the expression of the stress-energy field $V_2(z)$ is different from $T_{1+\infty}(z)$ in \cite{pro1411,Pro1}.

To obtain $V_3(z)$, we calculate
\beaa
U_1(z)U_3(w)&\sim &-\frac{1}{h_1h_2}\left(\frac{N(N-1)(N-2)\alpha_0^2}{(z-w)^4}+\frac{(N-1)(N-2)\alpha_0 U_1(w)}{(z-w)^3}+\frac{(N-2)U_2(w)}{(z-w)^2}\right),\\
U_1(z)U_1U_2(w)&\sim &-\frac{1}{h_1h_2}\left(\frac{N(N-1)\alpha_0U_1(w)}{(z-w)^3}+\frac{(N-1) U_1U_1(w)}{(z-w)^2}+\frac{NU_2(w)}{(z-w)^2}\right),\\
U_1(z)U_1U_1U_1(w)&\sim &-\frac{1}{h_1h_2}\frac{3NU_1U_1(w)}{(z-w)^2},\\
U_1(z)U_2^\prime(w)&\sim &-\frac{1}{h_1h_2}\left(\frac{3N(N-1)\alpha_0}{(z-w)^4}+\frac{2(N-1) U_1(w)}{(z-w)^3}+\frac{(N-1)U_1^\prime(w)}{(z-w)^2}\right),\\
U_1(z)U_1^{\prime\prime}(w)&\sim &-\frac{1}{h_1h_2}\frac{6N}{(z-w)^4},\\
U_1(z)U_1^\prime U_1(w)&\sim &-\frac{1}{h_1h_2}\left(\frac{2N U_1(w)}{(z-w)^3}+\frac{NU_1^\prime(w)}{(z-w)^2}\right),
\eeaa
Define
\bea
V_3(z)&=&h_1^2h_2^2\left(-U_3(z)+U_1U_2(z)-\frac{1}{3}U_1U_1U_1(z)+\frac{(N-2)\alpha_0}{2}U_2^\prime(z)\right.\nn\\
&&-\frac{\alpha_0^2(N-1)(N-2)}{12}U_1^{\prime\prime}(z)\left.-\frac{(N-1)\alpha_0}{2}U_1^\prime U_1(z)\right),\label{v3zU}
\eea
which equals
\bea
V_3(z)&=&-\frac{1}{3}h_1^2h_2^2\sum_{j=1}^N:J_1(z)^3:+\frac{1}{2}\alpha_0h_1^2h_2^2\sum_{j<k}J_jJ_k^\prime(z)-\frac{1}{2}\alpha_0h_1^2h_2^2\sum_{j<k}J_j^\prime J_k(z)\nn\\
&&-\frac{1}{2}\alpha_0h_1^2h_2^2\sum_{j=1}^N(N+1-2j)J_j^\prime J_j(z)\nn\\
&&+\alpha_0^2h_1^2h_2^2\sum_{j=1}^N\left(\frac{(j-1)(N-j)}{2}-\frac{(N-1)(N-j)}{12}\right)J_j^{\prime\prime}(z),
\eea
we have
\bea
V_1(z)V_3(w)&\sim&\frac{-2V_2(w)}{(z-w)^2}.
\eea
Note that here if we replace $V_3(w)$ by $-V_3(w)$, the minus before $V_2(w)$ in the OPE above will disappear. We use the expression of $V_3(z)$ in (\ref{v3zU}) since we want the coefficient of $U_3(z)$ in $V_3(z)$ to be $-1$, which matches the calculation in \cite{pro1411}. The OPEs
\bea
V_2(z)V_3(w)&\sim&\frac{\psi_0(\psi_0\sigma_2+\psi_0^3\sigma_3^2)V_1(w)}{(z-w)^4}+\frac{3V_3(w)}{(z-w)^2}+\frac{V_3^\prime(w)}{(z-w)}
\eea
and
\beaa
V_3(z)V_3(w)&\sim&\frac{c_3}{(z-w)^3}+\frac{\psi_0(4N+(N+2)N(N-2)\alpha_0^2h_1h_2)V_2(w)+\frac{3}{2}Nh_3\sigma_3U_1U_1(w)}{(z-w)^4}\nn\\
&&+\frac{\psi_0(4N+(N+2)N(N-2)\alpha_0^2h_1h_2)V_2^\prime(w)+{3}Nh_3\sigma_3U_1^\prime U_1(w)}{2(z-w)^3}+\cdots.
\eeaa
The central charge $c_3$
\beaa
c_3&=&\frac{1}{6\psi_0}(N+(N+1)N(N-1)\alpha_0^2h_1h_2)(4N+(N+2)N(N-2)\alpha_0^2h_1h_2)-\frac{N^2}{2}h_3^2,
\eeaa
where
\beaa
N+(N+1)N(N-1)\alpha_0^2h_1h_2=1-(1+\psi_0h_1h_2)(1+\psi_0h_1h_3)(1+\psi_0h_2h_3),\\
4N+(N+2)N(N-2)\alpha_0^2h_1h_2=8-(2+\psi_0h_1h_2)(2+\psi_0h_1h_3)(2+\psi_0h_2h_3).
\eeaa
\subsection{The $W_{1+\infty}$ algebra $\bar{V}_n(z)$}
From (\ref{ajnakmcom}) and (\ref{j(z)}), we know that
\be
J(z)J(w)\sim -\frac{N}{h_1h_2}\frac{1}{(z-w)^2}=\psi_0\frac{1}{(z-w)^2}.
\ee

Define
\bea
\bar{V}_1(z)&=&\frac{1}{\sqrt{\psi_0}}J(z),\\
\bar{V}_2(z)&=&\frac{1}{2\psi_0}:J(z)^2:,\\
\bar{V}_3(z)&=&\frac{1}{3\psi_0^{3/2}}:J(z)^3:,\\
\bar{V}_4(z)&=&\frac{1}{4\psi_0^2}:J(z)^4:-\frac{3}{20\psi_0}:J'(z)^2:+\frac{1}{10\psi_0}:J''(z)J(z):.
\eea
The center $\bar{c}_1$ of $\bar{V}_1(z)$ is $1$.
\bea
\bar{V}_1(z)\bar{V}_2(w)\sim \frac{\bar{V}_1(w)}{(z-w)^2}+\frac{\bar{V}_1^\prime(w)}{(z-w)},
\eea
and
\bea
\bar{V}_2(z)\bar{V}_2(w)\sim \frac{1/2}{(z-w)^4}+\frac{2\bar{V}_2(w)}{(z-w)^2}+\frac{\bar{V}_2^\prime(w)}{(z-w)},
\eea
which means the center $\bar{c}_2=1$. The OPEs related to $\bar{V}_3(z)$ are
\bea
\bar{V}_1(z)\bar{V}_3(w)&\sim &\frac{2\bar{V}_2(w)}{(z-w)^2},\\
\bar{V}_2(z)\bar{V}_3(w)&\sim &\frac{\bar{V}_1(w)}{(z-w)^4}+\frac{3\bar{V}_3(w)}{(z-w)^2}+\frac{\bar{V}_3^\prime(w)}{(z-w)},\\
\bar{V}_3(z)\bar{V}_3(w)&\sim &\frac{2/3}{(z-w)^6}+\frac{4\bar{V}_2(w)}{(z-w)^4}+\frac{2\bar{V}_2^\prime(w)}{(z-w)^3}\nn\\
&&+\frac{4\bar{V}_4(w)+\frac{3}{5}\bar{V}_2^{\prime\prime}(w)}{(z-w)^2}+\frac{2\bar{V}_4^\prime(w)+\frac{2}{15}\bar{V}_2^{\prime\prime\prime}(w)}{(z-w)}.
\eea
The OPEs related to $\bar{V}_4(z)$ are
\bea
\bar{V}_1(z)\bar{V}_4(w)&\sim &\frac{3}{5}\frac{\bar{V}_1(w)}{(z-w)^4}-\frac{3}{5}\frac{\bar{V}_1^\prime(w)}{(z-w)^3}+\left(3\bar{V}_3(w)+\frac{1}{10}\bar{V}_1^{\prime\prime}(w)\right)\frac{1}{(z-w)^2},\\
\bar{V}_2(z)\bar{V}_4(w)&\sim &\frac{21}{5}\frac{\bar{V}_2(w)}{(z-w)^4}+\frac{4\bar{V}_4(w)}{(z-w)^2}+\frac{\bar{V}_4^\prime(w)}{(z-w)},\\
\bar{V}_3(z)\bar{V}_4(w)&\sim &\frac{2\bar{V}_1(w)}{(z-w)^6}+\frac{54}{5}\frac{\bar{V}_3(w)}{(z-w)^4}+\frac{18}{5}\frac{\bar{V}_3^\prime(w)}{(z-w)^3}\nn\\
&&+\left(5\bar{V}_5(w)+a\bar{V}_3^{\prime\prime}(w)\right)\frac{1}{(z-w)^2}+\frac{2\bar{V}_5^\prime(w)+b\bar{V}_2^{\prime\prime\prime}(w)}{(z-w)},\\
\bar{V}_4(z)\bar{V}_4(w)&\sim &\frac{9}{5}\frac{1}{(z-w)^8}+\frac{72}{5}\frac{\bar{V}_2(w)}{(z-w)^6}+\frac{36}{5}\frac{\bar{V}_2^\prime(w)}{(z-w)^5}\nn\\
&&+\frac{1}{(z-w)^4}\left(\frac{114}{5}\bar{V}_4(w)+\frac{108}{50}\bar{V}_2^{\prime\prime}\right)+\frac{1}{(z-w)^3}\left(\frac{57}{5}\bar{V}_4^\prime(w)+\frac{12}{25}\bar{V}_2^{\prime\prime\prime}\right)\nn\\
&&+\frac{1}{(z-w)^2}\left(6\bar{V}_6(w)+\cdots\right)+\frac{1}{(z-w)}\left(\bar{V}_6^\prime(w)+\cdots\right).
\eea
Note that when $N=1$, $V_1(z)=\sqrt{\psi_0}\bar{V}_1(z),$ $V_2(z)=\bar{V}_2(z),$ $V_3(z)=3\sqrt{\psi_0}\bar{V}_3(z)$.

\section{3D Bosons}\label{sect4}
In this section, we show the 3D Boson fields and 3D Boson representation of $W_{1+\infty}$ algebra.
 \subsection{3D Boson fields}
 We denote 3D Boson fields by $B_j(z)$. Clearly, $B_1(z)=J(z)$. We found that $B_2(z)$ can not be $V_2(z)$ defined in (\ref{v2z}) since the central charge of $V_2(z)$ is $\psi_0\sigma_2+\psi_0^3\sigma_3^2$ which does not become zero when $h_1=h,h_2=-1/h, \psi_0=1$. We define
\be
B_2(z)=-2h_1h_2\left( -U_2(z) + \frac{(N-1)\alpha_0}{2} U_1^\prime(z) + \frac{N-1}{2N} (U_1 U_1)(z)\right),
\ee
which equals
\beaa
B_2(z)&=&-{h_1h_2}(1-\frac{1}{N})\sum_{j=1}^N :J_j(z)J_j(z):+\frac{2h_1h_2}{N}\sum_{j<k} :J_j(z)J_k(z):\\
&&-{\alpha_0h_1h_2}\sum_{j=1}^N(N+1-2j) J_j^\prime(z),
\eeaa
and
 satisfy
\beaa
B_1(z)B_2(w)&\sim &0,\\
B_2(z)B_2(w)&\sim &\frac{c_2^B}{(z-w)^4}+\frac{4B_2(w)}{(z-w)^2}+\frac{2B_2^\prime(w)}{(z-w)}
\eeaa
with the central charge
\[
c_2^B=-2(1+\psi_0\sigma_2+\psi_0^3\sigma_3^2),
\]
which equals $\langle P_{2,2},P_{2,2}\rangle$ in \cite{3-schur}. We notice that
\be
B_2(z)=2V_2(z)-2\bar{V}_2(z)
\ee
with
\be
V_2(z)\bar{V}_2(w)\sim \frac{1/2}{(z-w)^4}+\frac{2\bar{V}_2(w)}{(z-w)^2}+\frac{\bar{V}_2^\prime(w)}{(z-w)}.
\ee
Note that when $N=1$, that is, 3D Young diagrams become 2D Young diagrams, the 3D Boson $B_1(z)$ becomes 2D Boson field $J_1(z)$, and the 3D Boson field $B_2(z)$ become zero.

We define
\bea
B_3(z)&=&6h_1^2h_2^2\left(-U_3(z)+\frac{N-2}{N}U_1U_2(z)-\frac{(N-1)(N-2)}{3N^2}U_1U_1U_1(z)\right.\nn\\
&&+\frac{(N-2)\alpha_0}{2}
U_2^\prime(z)-\frac{(N-1)(N-2)\alpha_0^2}{12}U_1^{\prime\prime}(z)\nn\\
&&\left.-\frac{(N-1)(N-2)\alpha_0}{2N}U_1^\prime U_1(z)\right),
\eea
which satisfies
\beaa
B_1(z)B_3(w)&\sim & 0,\\
B_2(z)B_3(w)&\sim & \frac{6B_3(w)}{(z-w)^2}+\frac{2B_3^\prime(w)}{(z-w)},\\
B_3(z)B_3(w)&\sim &\frac{c_3^B}{(z-w)^6}+\frac{2(N-1)(2+\psi_0h_1h_2)(2+\psi_0h_1h_3)(2+\psi_0h_2h_3)B_2(w)}{\psi_0(z-w)^4}\\
&&+\frac{(N-1)(2+\psi_0h_1h_2)(2+\psi_0h_1h_3)(2+\psi_0h_2h_3)B_2^\prime(w)}{\psi_0(z-w)^3}+\cdots,
\eeaa
where
\beaa
c_3^B&=&\frac{6}{\psi_0}(N-1)(1+(N+1)N\alpha_0^2h_1h_2)(N-2)(4+(N+2)N\alpha_0^2h_1h_2)\nn\\
&=&\frac{6}{\psi_0}(1+\psi_0h_1h_2)(1+\psi_0h_1h_3)(1+\psi_0h_2h_3)(2+\psi_0h_1h_2)(2+\psi_0h_1h_3)(2+\psi_0h_2h_3).
\eeaa

From the calculations above, we see that the 3D Boson fields are Boson field $U_1(z)$ and the algebra $W_\infty$ which is given in \cite{pro1411}. We found that when $N=1$, the 3D Boson fields become 2D Boson field $U_1(z)=J_1(z)$, which corresponds to that 3D Young diagrams become 2D Young diagrams, concretely, when $N=1$, the 3D Boson field $B_1(z)$ becomes 2D Boson field $U_1(z)=J_1(z)$, and the 3D Boson fields $B_{j\geq 2}(z)$ become zero. These results match that when $N=1$, the 3D variables $P_{n,j\geq 2}$ become zero and 3-Jack polynomials become 2D symmetric functions $Y_\lambda$ (2D Jack polynomials when $h_1=h,h_2=-h^{-1}$)\cite{3-schur}.

\subsection{3D Boson representation of the $W_{1+\infty}$ algebra}
In this subsection, we use 3D Boson fields to represent the fields $V_n(z)$ in $W_{1+\infty}$ algebra. We see that $V_1(z)=B_1(z)$.
For $n=2$, 
\be
V_{2}(z)=\frac{1}{2}B_2(z)+\frac{1}{2\psi_0}B_1B_1(z),
\ee
which can also be written as
\be
B_2(z)=2V_2(z)-2\bar{V}_2(z).
\ee
When $n=3$,
\beaa
V_3(z)&=&\frac{1}{6}B_3(z)-\frac{2}{\psi_0}V_1V_2(z)+\frac{2}{3\psi_0^2}V_1V_1V_1(z),
\eeaa
which means
\bea
B_3(z)=6V_3(z)+\frac{12}{\psi_0}V_1V_2(z)-\frac{12}{\psi_0^{1/2}}\bar{V}_3(z).
\eea
We also have
\be
V_3(z)=\frac{1}{6}B_3(z)+\frac{1}{\psi_0}B_1B_2(z)+\frac{1}{3\psi_0^2}B_1B_1B_1(z).
\ee
\section{The Littlewood-Richardson rule for 3-Jack polynomials}\label{sect5}
In this section, we show the Littlewood-Richardson rule for 3-Jack polynomials.
Define
\be
B_j(z)=\sum_{n\in\Z} b_{n,j}z^{-n-j},
\ee
then
\beaa
b_{n,1}&=&\sum_{j=1}^Na_{j,n},\\
b_{n,2}&=&-h_1h_2(1-\frac{1}{N})\sum_{j=1}^N\sum_{k+l=n}:a_{j,k}a_{j,l}:+\frac{2h_1h_2}{N}\sum_{j<k}\sum_{m+l=n}:a_{j,m}a_{k,l}:\\
&&+h_3\sum_{j=1}^N(N+1-2j)(-n-1)a_{j,n},\\
b_{n,3}&=&6h_1^2h_2^2\left(-\sum_{j<k<l}\sum_{m+p+q=n}:a_{j,m}a_{k,p}a_{l,q}:-\alpha_0\sum_{j<k}\sum_{m+q=n}(j-1)(-m-1):a_{j,m}a_{k,q}:\right.\\
&&-\alpha_0\sum_{j<k}\sum_{m+q=n}(k-2)(-q-1):a_{j,m}a_{k,q}:-\frac{\alpha_0^2}{2}\sum_{j=1}^N(j-1)(j-2)(-n-1)(-n-2)a_{j,n}\\
&&+\frac{N-2}{N}\sum_{j=1}^N\sum_{k<l}\sum_{m+p+q=n}:a_{j,m}a_{k,p}a_{l,q}:+\frac{(N-2)\alpha_0}{N}\sum_{j,k=1}^N\sum_{m+q=n}(k-1)(-q-1):a_{j,m}a_{k,q}:\\
&&-\frac{(N-1)(N-2)}{3N^2}\sum_{j,k,l=1}^N\sum_{m+p+q=n}:a_{j,m}a_{k,p}a_{l,q}:+\frac{(N-2)\alpha_0^2}{2}\sum_{j=1}^N(j-1)(-n-1)(-n-2)a_{j,n}\\
&&+\frac{(N-2)\alpha_0}{2}\sum_{j<k}\sum_{m+q=n}(-m-q-2):a_{j,m}a_{k,q}:-\frac{(N-1)(N-2)\alpha_0^2}{12}\sum_{j=1}^N(-n-1)(-n-2)a_{j,n}\\
&&\left.-\frac{(N-1)(N-2)\alpha_0}{2N}\sum_{j,k=1}^N\sum_{m+q=n}(-m-1):a_{j,m}a_{k,q}:\right).
\eeaa

For $j=1,2,\cdots,N$, the Boson algebra $a_{j,n}$ has an representation on 2D Young diagrams $\lambda$ or the symmetric functions $Y_\lambda$, where $Y_\lambda$ are defined in \cite{WBCW} and are functions of power sums $p_n, n=1,2,\cdots$. The symmetric functions $Y_\lambda$ become 2D Jack polynomials when $h_1=h,h_2=-h^{-1}$. We denote the power sums related to $a_{j,n}$ by $p_{j,n}$. Denote the vacuum state by $|0\rangle_j$, the actions of $a_{j,n}$ on $|0\rangle_j$ are
\bea
a_{j,-n}|0\rangle_j=p_{j,n}|0\rangle_j,\ \ a_{j,n}|0\rangle_j=-\frac{1}{h_1h_2}\frac{\partial}{\partial p_{j,n}}|0\rangle_j,\ \ n>0.
\eea
Define
\be
|0\rangle=\sum_{j=1}^N|0\rangle_j,
\ee
3D Bosons are the operators acting on $|0\rangle$.
For $n>0$, define
\be
P_{n,k}=b_{-n,k}|0\rangle.
\ee
We calculate $P_{n,k}$ in order to obtain its property. When $k=1$,
\be
P_{n,1}=\sum_{j=1}^N a_{j,-n}|0\rangle=\sum_{j=1}^Np_{j,n}.
\ee
When $k=2$,
\bea
b_{-1,2}=-2h_1h_2(1-\frac{1}{N})\sum_{j=1}^N\sum_{l>0}a_{j,-l-1}a_{j,l}+\frac{2h_1h_2}{N}\sum_{j<k}\sum_{l>0}(a_{j,-l-1}a_{k,l}+a_{k,-l-1}a_{j,l}),
\eea
then $P_{1,2}=0$.
\bea
b_{-2,2}&=&-h_1h_2(1-\frac{1}{N})\sum_{j=1}^N(a_{j,-1}^2+2\sum_{l>0}a_{j,-l-2}a_{j,l})\nn\\
&&+\frac{2h_1h_2}{N}\sum_{j<k}\left(a_{j,-1}a_{k,-1}+\sum_{l>0}(a_{j,-l-2}a_{k,l}+a_{k,-l-2}a_{j,l})\right)\nn\\
&&-h_3\sum_{j=1}^N(N+1-2j)a_{j,-2},
\eea
then
\be
P_{2,2}=-h_1h_2(1-\frac{1}{N})\sum_{j=1}^Np_{j,1}^2+\frac{2h_1h_2}{N}\sum_{j<k}p_{j,1}p_{k,1}-h_3\sum_{j=1}^N(N+1-2j)p_{j,2},
\ee
which equals $P_{2,2}$ calculated in \cite{3-jack}. Generally,
\bea
P_{n,2}&=&-h_1h_2(1-\frac{1}{N})\sum_{j=1}^N\sum_{k,l>0,k+l=n}p_{j,k}p_{j,l}+\frac{2h_1h_2}{N}\sum_{j<k}\sum_{q,l>0,l+q=n}p_{j,l}p_{k,q}\nn\\
&&-h_3\sum_{j=1}^N(N+1-2j)(n-1)p_{j,n}.
\eea
When $k=3$,
\bea
P_{1,3}&=& 0,\\
P_{2,3}&=& 0,\\
P_{3,3}&=&6h_1^2h_2^2\left(-\sum_{j<k<l}p_{j,1}p_{k,1}p_{l,1}-\alpha_0\sum_{j<k}(j-1)p_{j,2}p_{k,1}\right.
-\alpha_0\sum_{j<k}(k-2)p_{j,1}p_{k,2}\nn\\
&&-{\alpha_0^2}\sum_{j=1}^N(j-1)(j-2)p_{j,3}+\frac{N-2}{N}\sum_{j=1}^N\sum_{k<l}p_{j,1}p_{k,1}p_{l,1}+\frac{(N-2)\alpha_0}{N}\sum_{j,k=1}^N(k-1)p_{j,1}p_{k,2}\nn\\
&&-\frac{(N-1)(N-2)}{3N^2}\sum_{j,k,l=1}^Np_{j,1}p_{k,1}p_{l,1}+{(N-2)\alpha_0^2}\sum_{j=1}^N(j-1)p_{j,3}\nn\\
&&+\frac{(N-2)\alpha_0}{2}\sum_{j<k}(p_{j,1}p_{k,2}+p_{j,2}p_{k,1})-\frac{(N-1)(N-2)\alpha_0^2}{6}\sum_{j=1}^Np_{j,3}\nn\\
&&\left.-\frac{(N-1)(N-2)\alpha_0}{2N}\sum_{j,k=1}^Np_{j,2}p_{k,1}\right),
\eea
and generally
\bea
P_{n,3}&=&6h_1^2h_2^2\left(-\sum_{j<k<l}\sum_{m+p+q=n}p_{j,m}p_{k,p}p_{l,q}-\alpha_0\sum_{j<k}\sum_{m+q=n}(j-1)(m-1)p_{j,m}p_{k,q}\right.\nn\\
&&-\alpha_0\sum_{j<k}\sum_{m+q=n}(k-2)(q-1)p_{j,m}p_{k,q}-\frac{\alpha_0^2}{2}\sum_{j=1}^N(j-1)(j-2)(n-1)(n-2)p_{j,n}\nn\\
&&+\frac{N-2}{N}\sum_{j=1}^N\sum_{k<l}\sum_{m+p+q=n}p_{j,m}p_{k,p}p_{l,q}+\frac{(N-2)\alpha_0}{N}\sum_{j,k=1}^N\sum_{m+q=n}(k-1)(q-1):p_{j,m}p_{k,q}\nn\\
&&-\frac{(N-1)(N-2)}{3N^2}\sum_{j,k,l=1}^N\sum_{m+p+q=n}p_{j,m}p_{k,p}p_{l,q}+\frac{(N-2)\alpha_0^2}{2}\sum_{j=1}^N(j-1)(n-1)(n-2)p_{j,n}\nn\\
&&+\frac{(N-2)\alpha_0}{2}\sum_{j<k}\sum_{m+q=n}(n-2)p_{j,m}p_{k,q}-\frac{(N-1)(N-2)\alpha_0^2}{12}\sum_{j=1}^N(n-1)(n-2)p_{j,n}\nn\\
&&\left.-\frac{(N-1)(N-2)\alpha_0}{2N}\sum_{j,k=1}^N\sum_{m+q=n}(m-1)p_{j,m}p_{k,q}\right),
\eea
where $j=1,2,\cdots, N$ and $m=1,2,\cdots$ in $p_{j,m}$.

We can see that $P_{n,j}=0$ when $n< j$, which means that 3-Jack polynomials are functions of variables $P_{n,j}$ with $n\geq j$. Define the degree of $P_{n,j}$ equal $n$, then the graded vector space of $P_{n,j}$ is isomorphic to that of 3D Young diagrams which is graded by the box numbers of 3D Young diagrams.

For 3D Young diagram $\begin{tikzpicture}
\draw (0,0)rectangle(0.25,0.25);
\draw (0,0.25) -- (0.175,0.35) [-];
\draw [shift = {+(0.25,0)}](0,0.25) -- (0.175,0.35) [-];
\draw [shift = {+(0.25,-0.250)}](0,0.25) -- (0.175,0.35) [-];
\draw (0.175,0.35) -- (0.425,0.35) [-];
\draw (0.425,0.35) -- (0.425,0.1) [-];
\end{tikzpicture}$, 3-Jack polynomial $\tilde{J}_{\begin{tikzpicture}
\draw (0,0)rectangle(0.25,0.25);
\draw (0,0.25) -- (0.175,0.35) [-];
\draw [shift = {+(0.25,0)}](0,0.25) -- (0.175,0.35) [-];
\draw [shift = {+(0.25,-0.250)}](0,0.25) -- (0.175,0.35) [-];
\draw (0.175,0.35) -- (0.425,0.35) [-];
\draw (0.425,0.35) -- (0.425,0.1) [-];
\end{tikzpicture}}$ equals
\[
\tilde{J}_{\Donebox}=P_{1,1}=\sum_{j=1}^Np_{j,1}.
\]
For 3D Young diagrams of two boxes, we have
\beaa
\tilde{J}_{\Dtwoboxy}&=&\frac{1}{(h_1-h_2)(h_1-h_3)}\left(\frac{1}{\psi_0}(1+h_2h_3\psi_0)P_{1,1}^2+(1+h_2h_3\psi_0)h_1P_{2,1}+P_{2,2}\right)\\
&=&  \frac{1}{\left ( h_1-h_2\right )\left ( h_1-h_3\right )}\left(-h_2\left (h_1-h_3\right ) \sum_{i=1}^Np_{i,1}^2+\sum_{i=1}^N\left(h_1-\right.\right.\nn\\
&&\left.\left.\left ( 2N-2i+1\right )h_3\right)p_{i,2}+2h_2h_3\sum_{i_1<i_2}p_{i_1,1}p_{i_2,1} \right),
\eeaa
where the first equation is equal to that in \cite{3-schur} when $\psi_0=1$, and the second equation is equal to that in \cite{3-jack}. Here what we want to do is constructing the Littlewood-Richardson rule for 3-Jack polynomials. We can see that the expressions of $P_{n,j}$ are much easier than that of $b_{-n,j}$. From \cite{3-jack}, we know the expressions of $P_{n,j}$, from them we can not obtain the expressions of $b_{-n,j}$. The 3D Bosons $b_{-n,j}$ are operators acting on 3D Young diagrams or 3-Jack polynomials, acting on the vacuum state $|0\rangle$,
\[
b_{-n,j}|0\rangle=P_{n,j}\cdot 1,
\]
where $\tilde{J}_0=1$, but for any 3D Young diagram $\pi$,
\[
b_{-n,j}|\pi\rangle\neq P_{n,j}\cdot \tilde{J}_\pi.
\]
This is the difference between 3D Bosons and 2D Bosons. For 2D Bosons $b_{-n,1}$ and any 2D Young diagram $\lambda$, we have
\[
b_{-n,1}|\lambda\rangle= P_{n,1}\cdot \tilde{J}_\lambda.
\]
Therefore for 2D Young diagrams or 2D Jack polynomials, in the Littlewood-Richardson rule
\be
\tilde{J}_\lambda(\{P_{n,1}\})\tilde{J}_\mu(\{P_{n,1}\})=\sum_{\nu}C_{\lambda\mu}^\nu\tilde{J}_\mu(\{P_{n,1}\}),
\ee
the 2D Jack polynomial $\tilde{J}_\mu(\{P_{n,1}\})$ is multiplied by $\tilde{J}_\lambda(\{P_{n,1}\})$, which equals $\tilde{J}_\lambda(\{b_{-n,1}\})$ acting on $\tilde{J}_\mu(\{P_{n,1}\})$.
For 3D Young diagrams or 3-Jack polynomials, the Littlewood-Richardson rule is given by
\be
\tilde{J}_\pi(\{P_{n,j}\})\times\tilde{J}_{\pi'}(\{P_{n,j}\}):=\tilde{J}_\pi(\{b_{-n,j}\})\cdot\tilde{J}_{\pi'}(\{P_{n,j}\}),
\ee
where in the left hand, $\times$ is not the multiplication in the ordinary way, and in the right hand side, $\cdot$ means the action, so $\tilde{J}_\pi(\{P_{n,j}\})\times\tilde{J}_{\pi'}(\{P_{n,j}\})$ is not equal to $\tilde{J}_{\pi'}(\{P_{n,j}\})$ multiplied by $\tilde{J}_\pi(\{P_{n,j}\})$.
\section{The representation of 3D Bosons on 3D Young diagrams}\label{sect6}
In this section, we show the representation of 3D Bosons on 3D Young diagrams or 3-Jack polynomials. Since 3-Jack polynomials are functions of variables $p_{j,n}$ with $j=1,2,\cdots,N$ and $n=1,2,\cdots$, the actions of 3D Bosons $b_{n,j}$ on 3-Jack polynomials $\tilde{J}_\pi$ can be determined by
\bea
a_{j,-n}\cdot \tilde{J}_\pi(\{p_{j,n}\}) &=&p_{j,n}\tilde{J}_\pi(\{p_{j,n}\}),\\
a_{j,n}\cdot \tilde{J}_\pi(\{p_{j,n}\}) &=&-\frac{1}{h_1h_2}n \frac{\partial}{\partial p_{j,n}}\tilde{J}_\pi(\{p_{j,n}\}),
\eea
for $n>0$.

Since we have known the representation of affine Yangian of ${\mathfrak{gl}}(1)$ on 3D Young diagrams, in the following, we use the generators of  affine Yangian of ${\mathfrak{gl}}(1)$ to represent 3D Bosons.
We know that \cite{WBCW}
\bea
\psi_2&=&-2h_1h_2\sum_{j=1}^N\sum_{k>0}a_{j,-k}a_{j,k},\\
\psi_3 &=& 3h_1^2h_2^2\sum_{i=1}^N\sum_{j,k>0}(a_{i,-j-k}a_{i,j}a_{i,k}
+a_{i,-j}a_{i,-k}a_{i,j+k})\nn\\
&&+6\sigma_3\sum_{i_1<i_2}\sum_{k>0}ka_{i_1,-k}a_{i_2,k}+(-4N+6j-3)\sigma_3\sum_{j=1}^N\sum_{k>0}a_{j,-k}a_{j,k}\nn\\
&&+3\sigma_3\sum_{j=1}^N\sum_{k>0}ka_{j,-k}a_{j,k},
\eea
and
\bea
e_0&=&\sum_{j=1}^N a_{j,-1},\\
e_1&=& -h_1h_2\sum_{j=1}^N\sum_{k>0}a_{j,-k-1}a_{j,k}.
\eea

Compare with the expression of $b_{n,1}$, we obtain that
\bea
b_{-n,1}&=&\frac{1}{(n-1)!}\text{ad}_{e_1}^{n-1}e_0,\label{b-n1e1e0}\\
b_{n,1}&=&-\frac{1}{(n-1)!}\text{ad}_{f_1}^{n-1}f_0\label{bn1e1e0}
\eea
for $n>0$, since
\[
[a_{j,-k-1}a_{j,k},a_{i,-l}]=-\frac{1}{h_1h_2}\delta_{i,j}\delta_{k,l}la_{j,-l-1}.
\]
The expressions in (\ref{b-n1e1e0}) and (\ref{bn1e1e0}) equals that in \cite{WBCW}.

Let $\psi_0=1$ without loss the generality. For Boson field $B_2(z)$, we have
\beaa
b_{-1,2}&=&e_1-2\sum_{n\geq 1}b_{-n-1,1}b_{n,1},\\
 b_{1,2}&=&-f_1-2\sum_{n\geq 1}b_{-n,1}b_{n+1,1},
\eeaa
which also explains $b_{-1,2}|0\rangle=0$ since $e_1|0\rangle=0$.
For $n\geq 1$,
\bea
b_{-(n+1),2}&=&\frac{1}{(n-1)!}\text{ad}_{e_1}^{n-1}\left([e_2,e_0]-\sigma_3[e_1,e_0]\right)-\sum_{i+j=-(n+1)}:b_{i,1}b_{j,1}:\\
b_{n+1,2}&=&-\frac{1}{(n-1)!}\text{ad}_{f_1}^{n-1}\left([f_2,f_0]-\sigma_3[f_1,f_0]\right)-\sum_{i+j=(n+1)}:b_{i,1}b_{j,1}:,
\eea
which equal that in \cite{3DbosonYangian}.

As operators,
\beaa
\tilde{J}_{\Dtwoboxy}&=&\frac{1}{(h_1-h_2)(h_1-h_3)}\left((1+h_2h_3)b_{-1,1}^2+(1+h_2h_3)h_1b_{-2,1}+b_{-2,2}\right)\\
&=&\frac{1}{(h_1-h_2)(h_1-h_3)}\left(h_2h_3e_0^2+h_1[e_1,e_0]+[e_2,e_0]-2\sum_{n\geq 1}b_{-n-2,1}b_{n,1}\right),\\
\eeaa
and
\beaa
\tilde{J}_{\Dtwoboxx}&=&\frac{1}{(h_2-h_1)(h_2-h_3)}\left(h_1h_3e_0^2+h_2[e_1,e_0]+[e_2,e_0]-2\sum_{n\geq 1}b_{-n-2,1}b_{n,1}\right).\\
\eeaa

From
\bea
e_j\cdot\tilde{J}_\pi&=&\sum_{\Box\in\pi^+} h_\Box^j \tilde{J}_{\pi+\Box},\\
f_j\cdot\tilde{J}_\pi&=&-\sum_{\Box\in\pi^-} h_\Box^j F^2(\pi\rightarrow \pi-\Box)\tilde{J}_{\pi+\Box},
\eea
we have
\[
b_{1,1}\cdot \tilde{J}_{\Dtwoboxx}= \frac{2(1+h_1h_3)}{(h_2-h_1)(h_2-h_3)}\tilde{J}_{\Donebox},\ \ \ b_{2,1}\cdot \tilde{J}_{\Dtwoboxy}= \frac{2(1+h_1h_3)h_2}{(h_2-h_1)(h_2-h_3)},
\]
and
\beaa
b_{-3,1}\cdot \tilde{J}_{\Donebox}&=&\frac{1}{2}\left(e_1e_1e_0-2e_1e_0e_1+e_0e_1e_1\right)\cdot \tilde{J}_{\Donebox}\\
&=& h_1^2\tilde{J}_{\scalebox{0.08}{\includegraphics{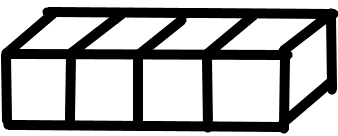}}}+  h_2^2\tilde{J}_{\scalebox{0.1}{\includegraphics{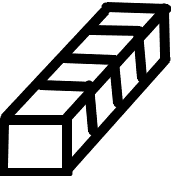}}} +h_3^2\tilde{J}_{\scalebox{0.1}{\includegraphics{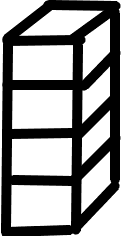}}}\\
&&+h_1^2\tilde{J}_{\scalebox{0.08}{\includegraphics{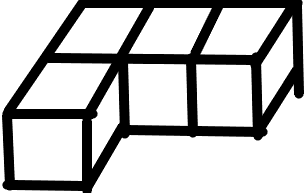}}_{h_1,2h_1,h_2}}+\frac{3h_1h_2-4h_1^2}{2}\tilde{J}_{\scalebox{0.08}{\includegraphics{3Dfourboxyyx.PNG}}_{h_1,h_2,2h_1}}+\frac{2h_1^2-3h_1h_2}{2}\tilde{J}_{\scalebox{0.08}{\includegraphics{3Dfourboxyyx.PNG}}_{h_2,h_1,2h_1}}\\
&&+h_1^2\tilde{J}_{\scalebox{0.08}{\includegraphics{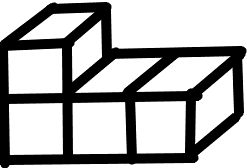}}_{h_1,2h_1,h_3}}+\frac{3h_1h_3-4h_1^2}{2}\tilde{J}_{\scalebox{0.08}{\includegraphics{3Dfourboxyyz.PNG}}_{h_1,h_3,2h_1}}+\frac{2h_1^2-3h_1h_3}{2}\tilde{J}_{\scalebox{0.08}{\includegraphics{3Dfourboxyyz.PNG}}_{h_3,h_1,2h_1}}\\
&&+h_2^2\tilde{J}_{\scalebox{0.08}{\includegraphics{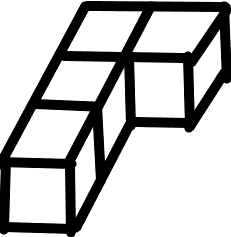}}_{h_2,2h_2,h_1}}+\frac{3h_1h_2-4h_2^2}{2}\tilde{J}_{\scalebox{0.08}{\includegraphics{3Dfourboxyxx.PNG}}_{h_2,h_1,2h_2}}+\frac{2h_2^2-3h_1h_2}{2}\tilde{J}_{\scalebox{0.08}{\includegraphics{3Dfourboxyxx.PNG}}_{h_1,h_2,2h_2}}\\
&&+h_2^2\tilde{J}_{\scalebox{0.08}{\includegraphics{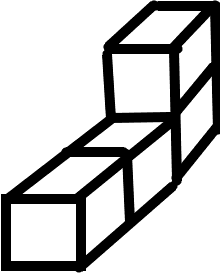}}_{h_2,2h_2,h_3}}+\frac{3h_2h_3-4h_2^2}{2}\tilde{J}_{\scalebox{0.08}{\includegraphics{3Dfourboxxxz.PNG}}_{h_2,h_3,2h_2}}+\frac{2h_2^2-3h_2h_3}{2}\tilde{J}_{\scalebox{0.08}{\includegraphics{3Dfourboxxxz.PNG}}_{h_3,h_2,2h_2}}\\
&&+h_3^2\tilde{J}_{\scalebox{0.08}{\includegraphics{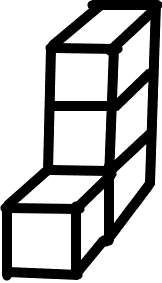}}_{h_3,2h_3,h_2}}+\frac{3h_2h_3-4h_3^2}{2}\tilde{J}_{\scalebox{0.08}{\includegraphics{3Dfourboxxzz.PNG}}_{h_3,h_2,2h_3}}+\frac{2h_3^2-3h_2h_3}{2}\tilde{J}_{\scalebox{0.08}{\includegraphics{3Dfourboxxzz.PNG}}_{h_2,h_3,2h_3}}\\
&&+h_3^2\tilde{J}_{\scalebox{0.1}{\includegraphics{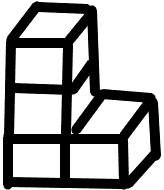}}_{h_3,2h_3,h_1}}+\frac{3h_1h_3-4h_3^2}{2}\tilde{J}_{\scalebox{0.1}{\includegraphics{3Dfourboxyzz.PNG}}_{h_3,h_1,2h_3}}+\frac{2h_3^2-3h_1h_3}{2}\tilde{J}_{\scalebox{0.1}{\includegraphics{3Dfourboxyzz.PNG}}_{h_1,h_3,2h_3}}\\
&&+\frac{h_3^2-2h_1^2}{2}\tilde{J}_{\scalebox{0.07}{\includegraphics{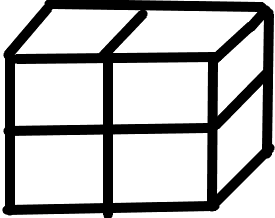}}_{h_1,h_3,h_1+h_3}}
+\frac{h_1^2-2h_3^2}{2}\tilde{J}_{\scalebox{0.07}{\includegraphics{3Dfourboxyzy.PNG}}_{h_3,h_1,h_1+h_3 }}\eeaa
\beaa
&&+\frac{h_2^2-2h_1^2}{2}\tilde{J}_{\scalebox{0.1}{\includegraphics{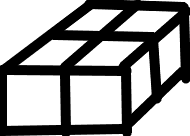}}_{h_1,h_2,h_1+h_2}}
+\frac{h_1^2-2h_2^2}{2}\tilde{J}_{\scalebox{0.1}{\includegraphics{3Dfourboxyxy.PNG}}_{h_2,h_1,h_1+h_2}}\\
&&+\frac{h_3^2-2h_2^2}{2}\tilde{J}_{\scalebox{0.07}{\includegraphics{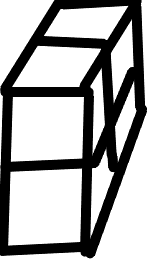}}_{h_2,h_3,h_2+h_3}}
+\frac{h_2^2-2h_3^2}{2}\tilde{J}_{\scalebox{0.07}{\includegraphics{3Dfourboxxzx.PNG}}_{h_3,h_2,h_2+h_3 }}\\
&&+\frac{-h_2^2-2h_1h_3}{2}\left(\tilde{J}_{\scalebox{0.08}{\includegraphics{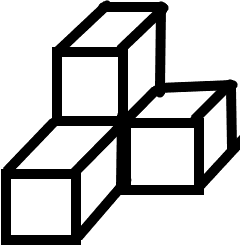}}_{h_3,h_2,h_1 }}+\tilde{J}_{\scalebox{0.08}{\includegraphics{3Dfourboxxyz.PNG}}_{h_1,h_2,h_3}}\right)\\
&&+\frac{-h_1^2-2h_2h_3}{2}\left(\tilde{J}_{\scalebox{0.08}{\includegraphics{3Dfourboxxyz.PNG}}_{h_2,h_1,h_3 }}+\tilde{J}_{\scalebox{0.08}{\includegraphics{3Dfourboxxyz.PNG}}_{h_3,h_1,h_2}}\right)\\
&&+\frac{-h_3^2-2h_1h_2}{2}\left(\tilde{J}_{\scalebox{0.08}{\includegraphics{3Dfourboxxyz.PNG}}_{h_1,h_3,h_2 }}+\tilde{J}_{\scalebox{0.08}{\includegraphics{3Dfourboxxyz.PNG}}_{h_2,h_3,h_1}}\right),
\eeaa
\beaa
b_{-4,1}\cdot \tilde{J}_{0}&=&\frac{1}{6}e_1e_1e_1\cdot \tilde{J}_{\Donebox}\\
&=& h_1^3\tilde{J}_{\scalebox{0.08}{\includegraphics{3Dfourboxy.PNG}}}+  h_2^3\tilde{J}_{\scalebox{0.1}{\includegraphics{3Dfourboxx.PNG}}} +h_3^3\tilde{J}_{\scalebox{0.1}{\includegraphics{3Dfourboxz.PNG}}}\\
&&+\frac{h_1^2h_2}{3}\left(\tilde{J}_{\scalebox{0.08}{\includegraphics{3Dfourboxyyx.PNG}}_{h_1,2h_1,h_2}}+\tilde{J}_{\scalebox{0.08}{\includegraphics{3Dfourboxyyx.PNG}}_{h_1,h_2,2h_1}}+\tilde{J}_{\scalebox{0.08}{\includegraphics{3Dfourboxyyx.PNG}}_{h_2,h_1,2h_1}}\right)\\
&&+\frac{h_1^2h_3}{3}\left(\tilde{J}_{\scalebox{0.08}{\includegraphics{3Dfourboxyyz.PNG}}_{h_1,2h_1,h_3}}+\tilde{J}_{\scalebox{0.08}{\includegraphics{3Dfourboxyyz.PNG}}_{h_1,h_3,2h_1}}+\tilde{J}_{\scalebox{0.08}{\includegraphics{3Dfourboxyyz.PNG}}_{h_3,h_1,2h_1}}\right)\\
&&+\frac{h_2^2h_1}{3}\left(\tilde{J}_{\scalebox{0.08}{\includegraphics{3Dfourboxyxx.PNG}}_{h_2,2h_2,h_1}}+\tilde{J}_{\scalebox{0.08}{\includegraphics{3Dfourboxyxx.PNG}}_{h_2,h_1,2h_2}}+\tilde{J}_{\scalebox{0.08}{\includegraphics{3Dfourboxyxx.PNG}}_{h_1,h_2,2h_2}}\right)\\
&&+\frac{h_2^2h_3}{3}\left(\tilde{J}_{\scalebox{0.08}{\includegraphics{3Dfourboxxxz.PNG}}_{h_2,2h_2,h_3}}+\tilde{J}_{\scalebox{0.08}{\includegraphics{3Dfourboxxxz.PNG}}_{h_2,h_3,2h_2}}+\tilde{J}_{\scalebox{0.08}{\includegraphics{3Dfourboxxxz.PNG}}_{h_3,h_2,2h_2}}\right)\\
&&+\frac{h_3^2h_2}{3}\left(\tilde{J}_{\scalebox{0.08}{\includegraphics{3Dfourboxxzz.PNG}}_{h_3,2h_3,h_2}}+\tilde{J}_{\scalebox{0.08}{\includegraphics{3Dfourboxxzz.PNG}}_{h_3,h_2,2h_3}}+\tilde{J}_{\scalebox{0.08}{\includegraphics{3Dfourboxxzz.PNG}}_{h_2,h_3,2h_3}}\right)\\
&&+\frac{h_3^2h_1}{3}\left(\tilde{J}_{\scalebox{0.1}{\includegraphics{3Dfourboxyzz.PNG}}_{h_3,2h_3,h_1}}+\tilde{J}_{\scalebox{0.1}{\includegraphics{3Dfourboxyzz.PNG}}_{h_3,h_1,2h_3}}+\tilde{J}_{\scalebox{0.1}{\includegraphics{3Dfourboxyzz.PNG}}_{h_1,h_3,2h_3}}\right)\\
&&+\frac{h_1h_3(h_1+h_3)}{6}\left(\tilde{J}_{\scalebox{0.07}{\includegraphics{3Dfourboxyzy.PNG}}_{h_1,h_3,h_1+h_3}}
+\tilde{J}_{\scalebox{0.07}{\includegraphics{3Dfourboxyzy.PNG}}_{h_3,h_1,h_1+h_3 }}\right)\\
&&+\frac{h_1h_2(h_1+h_2)}{6}\left(\tilde{J}_{\scalebox{0.1}{\includegraphics{3Dfourboxyxy.PNG}}_{h_1,h_2,h_1+h_2}}
+\tilde{J}_{\scalebox{0.1}{\includegraphics{3Dfourboxyxy.PNG}}_{h_2,h_1,h_1+h_2}}\right)\\
&&+\frac{h_2h_3(h_2+h_3)}{6}\left(\tilde{J}_{\scalebox{0.07}{\includegraphics{3Dfourboxxzx.PNG}}_{h_2,h_3,h_2+h_3}}
+\tilde{J}_{\scalebox{0.07}{\includegraphics{3Dfourboxxzx.PNG}}_{h_3,h_2,h_2+h_3 }}\right)\\
&&+\frac{h_1h_2h_3}{6}\left(\tilde{J}_{\scalebox{0.08}{\includegraphics{3Dfourboxxyz.PNG}}_{h_3,h_2,h_1 }}+\tilde{J}_{\scalebox{0.08}{\includegraphics{3Dfourboxxyz.PNG}}_{h_1,h_2,h_3}}+\tilde{J}_{\scalebox{0.08}{\includegraphics{3Dfourboxxyz.PNG}}_{h_2,h_1,h_3 }}+\tilde{J}_{\scalebox{0.08}{\includegraphics{3Dfourboxxyz.PNG}}_{h_3,h_1,h_2}}\right.\\
&&\left.+\tilde{J}_{\scalebox{0.08}{\includegraphics{3Dfourboxxyz.PNG}}_{h_1,h_3,h_2 }}+\tilde{J}_{\scalebox{0.08}{\includegraphics{3Dfourboxxyz.PNG}}_{h_2,h_3,h_1}}\right),
\eeaa
we calculate
\beaa
&&\tilde{J}_{\Dtwoboxy}(\{P_{n,j}\})\times \tilde{J}_{\Dtwoboxx}(\{P_{n,j}\})=\tilde{J}_{\Dtwoboxy}(\{b_{-n,j}\})\cdot \tilde{J}_{\Dtwoboxx}(\{P_{n,j}\})\\
&=&\frac{1}{(h_1-h_2)(h_1-h_3)}\left(h_2h_3e_0^2+h_1[e_1,e_0]+[e_2,e_0]-2(b_{-3,1}b_{1,1}+b_{-4,1}b_{2,1})\right)\cdot \tilde{J}_{\Dtwoboxx}\\
&=&\frac{-4h_1^2(1+h_1h_3)(1+h_1h_2)}{(h_1-h_2)(h_1-h_3)(h_2-h_1)(h_2-h_3)}\tilde{J}_{\scalebox{0.08}{\includegraphics{3Dfourboxy.PNG}}}+
\frac{-4h_3^2(1+h_1h_3)(1+h_2h_3)}{(h_1-h_2)(h_1-h_3)(h_2-h_1)(h_2-h_3)}\tilde{J}_{\scalebox{0.08}{\includegraphics{3Dfourboxz.PNG}}}\\
&&+\frac{-4h_2^2(1+h_2h_3)(1+h_1h_2)}{(h_1-h_2)(h_1-h_3)(h_2-h_1)(h_2-h_3)}\tilde{J}_{\scalebox{0.1}{\includegraphics{3Dfourboxx.PNG}}}
+\frac{-4h_1^2(1+h_1h_3)(3+h_2^2)}{3(h_1-h_2)(h_1-h_3)(h_2-h_1)(h_2-h_3)}\tilde{J}_{\scalebox{0.08}{\includegraphics{3Dfourboxyyx.PNG}}_{h_1,2h_1,h_2}}\\
&&+\frac{-2h_1^2(1+h_1h_3)(9h_1h_2-12h_1^2+2h_1^2h_2^2)}{3(h_1-h_2)(h_1-h_3)(h_2-h_1)(h_2-h_3)}\tilde{J}_{\scalebox{0.08}{\includegraphics{3Dfourboxyyx.PNG}}_{h_1,h_2,2h_1}}\\
&&+\frac{4h_1^4h_2^2+4h_1^3h_2^3-15h_1^3h_2+8h_1^2h_2^2-3h_1h_2^3-6h_2^4-12h_1^2+18h_1h_2}{3(h_1-h_2)(h_1-h_3)(h_2-h_1)(h_2-h_3)}\tilde{J}_{\scalebox{0.08}{\includegraphics{3Dfourboxyyx.PNG}}_{h_2,h_1,2h_1}}\\
&&+\frac{-4h_1^2(1+h_1h_3)(3+h_2h_3)}{3(h_1-h_2)(h_1-h_3)(h_2-h_1)(h_2-h_3)}\tilde{J}_{\scalebox{0.08}{\includegraphics{3Dfourboxyyz.PNG}}_{h_1,2h_1,h_3}}\\
&&+\frac{-2(1+h_1h_3)(9h_1h_3-12h_1^2+2h_1^2h_2h_3)}{3(h_1-h_2)(h_1-h_3)(h_2-h_1)(h_2-h_3)}\tilde{J}_{\scalebox{0.08}{\includegraphics{3Dfourboxyyz.PNG}}_{h_1,h_3,2h_1}}\\
&&+\frac{-2(1+h_1h_3)(6h_1^2-9h_1h_3+2h_1^2h_2h_3)}{3(h_1-h_2)(h_1-h_3)(h_2-h_1)(h_2-h_3)}\tilde{J}_{\scalebox{0.08}{\includegraphics{3Dfourboxyyz.PNG}}_{h_3,h_1,2h_1}}
\eeaa
\beaa
&&+\frac{4h_1^3h_2^3+4h_1^2h_2^4-6h_1^4+3h_1^3h_2+48h_1^2h_2^2+5h_1h_2^3-30h_2^4-12h_2^2}{3(h_1-h_2)(h_1-h_3)(h_2-h_1)(h_2-h_3)}\tilde{J}_{\scalebox{0.08}{\includegraphics{3Dfourboxyxx.PNG}}_{h_2,2h_2,h_1}}\\
&&+\frac{4h_1^3h_2^3+4h_1^2h_2^4+6h_1^4+21h_1^3h_2-30h_1^2h_2^2-31h_1h_2^3+18h_2^4-18h_1h_2+24h_2^2}{3(h_1-h_2)(h_1-h_3)(h_2-h_1)(h_2-h_3)}\tilde{J}_{\scalebox{0.08}{\includegraphics{3Dfourboxyxx.PNG}}_{h_2,2h_2,h_1}}\\
&&+\frac{-(2(h_1h_3+1))h_2(2h_1h_2^2-9h_1+6h_2)}{3(h_1-h_2)(h_1-h_3)(h_2-h_1)(h_2-h_3)}\tilde{J}_{\scalebox{0.08}{\includegraphics{3Dfourboxyxx.PNG}}_{h_1,h_2,2h_2}}\\
&&+\frac{-4h_1^3h_2^3-8h_1^2h_2^4-4h_1h_2^5+6h_1^3h_2+30h_1^2h_2^2+16h_1h_2^3-20h_2^4-12h_2^2}{3(h_1-h_2)(h_1-h_3)(h_2-h_1)(h_2-h_3)}\tilde{J}_{\scalebox{0.08}{\includegraphics{3Dfourboxxxz.PNG}}_{h_2,2h_2,h_3}}\\
&&+\frac{-4h_1^3h_2^3-8h_1^2h_2^4-4h_1h_2^5-18h_1^3h_2-66h_1^2h_2^2-44h_1h_2^3+16h_2^4+18h_1h_2+42h_2^2}{3(h_1-h_2)(h_1-h_3)(h_2-h_1)(h_2-h_3)}\tilde{J}_{\scalebox{0.08}{\includegraphics{3Dfourboxxxz.PNG}}_{h_2,h_3,2h_2}}\\
&&+\frac{-(2(h_1h_3+1))h_2(2h_2^2h_3+6h_2-9h_3)}{3(h_1-h_2)(h_1-h_3)(h_2-h_1)(h_2-h_3)}\tilde{J}_{\scalebox{0.08}{\includegraphics{3Dfourboxxxz.PNG}}_{h_2,h_3,2h_2}}
\eeaa
\beaa
&&+\frac{-(4(h_1h_3+1))h_3^2(h_2^2+3)}{3(h_1-h_2)(h_1-h_3)(h_2-h_1)(h_2-h_3)}\tilde{J}_{\scalebox{0.08}{\includegraphics{3Dfourboxxzz.PNG}}_{h_3,2h_3,h_2}}\\
&&+\frac{-(2(h_1h_3+1))h_3(2h_2^2h_3+9h_2-12h_3)}{3(h_1-h_2)(h_1-h_3)(h_2-h_1)(h_2-h_3)}\tilde{J}_{\scalebox{0.08}{\includegraphics{3Dfourboxxzz.PNG}}_{h_3,h_2,2h_3}}\\
&&+\frac{-2h_3(2h_1h_2^2h_3^2-6h_1h_2h_3+3h_1h_3^2-h_2^2h_3+3h_2h_3^2-9h_2+6h_3)}{3(h_1-h_2)(h_1-h_3)(h_2-h_1)(h_2-h_3)}\tilde{J}_{\scalebox{0.08}{\includegraphics{3Dfourboxxzz.PNG}}_{h_2,h_3,2h_3}}\\
&&+\frac{-(4(h_1h_3+1))h_3^2(h_1h_2+3)}{3(h_1-h_2)(h_1-h_3)(h_2-h_1)(h_2-h_3)}\tilde{J}_{\scalebox{0.1}{\includegraphics{3Dfourboxyzz.PNG}}_{h_3,2h_3,h_1}}\\
&&+\frac{-(2(h_1h_3+1))h_3(2h_1h_2h_3+9h_1-12h_3)}{3(h_1-h_2)(h_1-h_3)(h_2-h_1)(h_2-h_3)}\tilde{J}_{\scalebox{0.1}{\includegraphics{3Dfourboxyzz.PNG}}_{h_3,h_1,2h_3}}\\
&&+\frac{-2(h_1h_3+1)h_3(2h_1h_2h_3-9h_1+6h_3)}{3(h_1-h_2)(h_1-h_3)(h_2-h_1)(h_2-h_3)}\tilde{J}_{\scalebox{0.1}{\includegraphics{3Dfourboxyzz.PNG}}_{h_1,h_3,2h_3}}\\
\eeaa
\beaa
&&+\frac{-(2(h_1h_3+1))(h_1^2h_3+h_1h_3^2-6h_1^2+3h_3^2)}{3(h_1-h_2)(h_1-h_3)(h_2-h_1)(h_2-h_3)}\tilde{J}_{\scalebox{0.1}{\includegraphics{3Dfourboxyzy.PNG}}_{h_1,h_3,h_1+h_3}}\\
&&+\frac{-(2(h_1h_3+1))(h_1^2h_3+h_1h_3^2+3h_1^2-6h_3^2)}{3(h_1-h_2)(h_1-h_3)(h_2-h_1)(h_2-h_3)}\tilde{J}_{\scalebox{0.1}{\includegraphics{3Dfourboxyzy.PNG}}_{h_3,h_1,h_1+h_3}}\\
&&+\frac{-(2(h_1h_3+1))(h_1^2h_2+h_1h_2^2-6h_1^2+3h_2^2)}{3(h_1-h_2)(h_1-h_3)(h_2-h_1)(h_2-h_3)}\tilde{J}_{\scalebox{0.1}{\includegraphics{3Dfourboxyxy.PNG}}_{h_1,h_2,h_1+h_2}}\\
&&+\frac{(-2h_3(h_1^3h_2+h_1^2h_2^2+3h_1^3-6h_1h_2^2-h_1h_2-3h_1+6h_2)}{3(h_1-h_2)(h_1-h_3)(h_2-h_1)(h_2-h_3)}\tilde{J}_{\scalebox{0.1}{\includegraphics{3Dfourboxyxy.PNG}}_{h_2,h_1,h_1+h_2}}\\
&&+\frac{2h_1(h_1^3h_2+2h_1^2h_2^2+h_1h_2^3+3h_1^3+9h_1^2h_2+3h_1h_2^2-3h_2^3-h_1h_2-h_2^2-3h_1-9h_2)}{3(h_1-h_2)(h_1-h_3)(h_2-h_1)(h_2-h_3)}\tilde{J}_{\scalebox{0.1}{\includegraphics{3Dfourboxxzx.PNG}}_{h_2,h_3,h_2+h_3}}\\
&&+\frac{-(2(h_1h_3+1))(h_2^2h_3+h_2h_3^2+3h_2^2-6h_3^2)}{3(h_1-h_2)(h_1-h_3)(h_2-h_1)(h_2-h_3)}\tilde{J}_{\scalebox{0.1}{\includegraphics{3Dfourboxxzx.PNG}}_{h_3,h_2,h_2+h_3}}
\eeaa
\beaa
&&+\frac{-(2(h_1h_3+1))(h_1h_2h_3-6h_1h_3-3h_2^2)}{3(h_1-h_2)(h_1-h_3)(h_2-h_1)(h_2-h_3)}\left(\tilde{J}_{\scalebox{0.1}{\includegraphics{3Dfourboxxyz.PNG}}_{h_1,h_2,h_3}}+\tilde{J}_{\scalebox{0.1}{\includegraphics{3Dfourboxxyz.PNG}}_{h_3,h_2,h_1}}\right)\\
&&+\frac{-(2(h_1h_3+1))(3h_1^2-h_3h_2+h_3h_2h_1)+3h_3h_2+3h_1h_3-6h_1^2+3h_3^2}{3(h_1-h_2)(h_1-h_3)(h_2-h_1)(h_2-h_3)}\tilde{J}_{\scalebox{0.1}{\includegraphics{3Dfourboxxyz.PNG}}_{h_2,h_1,h_3}}\\
&&+\frac{-(2(h_1h_3+1))(h_1h_2h_3+3h_1^2-6h_2h_3)}{3(h_1-h_2)(h_1-h_3)(h_2-h_1)(h_2-h_3)}\tilde{J}_{\scalebox{0.1}{\includegraphics{3Dfourboxxyz.PNG}}_{h_3,h_1,h_2}}\\
&&+\frac{-(2(h_1h_3+1))(h_1h_2h_3-6h_1h_2+3h_3^2)}{3(h_1-h_2)(h_1-h_3)(h_2-h_1)(h_2-h_3)}\tilde{J}_{\scalebox{0.1}{\includegraphics{3Dfourboxxyz.PNG}}_{h_1,h_3,h_2}}\\
&&+\frac{-(2(h_1h_3+1))(-6h_1h_2+3h_3^2+h_3h_2h_1)+3h_3h_2-3h_1h_3+6h_1^2-3h_3^2}{3(h_1-h_2)(h_1-h_3)(h_2-h_1)(h_2-h_3)}\tilde{J}_{\scalebox{0.1}{\includegraphics{3Dfourboxxyz.PNG}}_{h_1,h_3,h_2}}.
\eeaa
This is the Littlewood-Richardson rule for $\tilde{J}_{\Dtwoboxy}(\{P_{n,j}\})\times \tilde{J}_{\Dtwoboxx}(\{P_{n,j}\})$. 
We can check that
\[
\tilde{J}_{\Dtwoboxy}(\{P_{n,j}\})\times \tilde{J}_{\Dtwoboxx}(\{P_{n,j}\})=\tilde{J}_{\Dtwoboxx}(\{P_{n,j}\})\times \tilde{J}_{\Dtwoboxy}(\{P_{n,j}\}),
\]
and 
\beaa
\tilde{J}_{\Dtwoboxy} \left(\tilde{J}_{\Dtwoboxx}+\tilde{J}_{\Dtwoboxy}+\tilde{J}_{\Dtwoboxz}\right)=\tilde{J}_{\Dtwoboxy}\tilde{J}_{\Donebox}^2=\tilde{J}_{\Donebox}^2\tilde{J}_{\Dtwoboxy}.
\eeaa
We can see that the Littlewood-Richardson rule for 3-Jack polynomials are complicated. When $h_1=1,h_2=-1$, the Littlewood-Richardson rule for 3-Jack polynomials becomes that for Schur functions: when $h_1=1,h_2=-1$, 3-Jack polynomials of 3D Young diagrams which have more than one layer in $z$-axis direction become zero, and 3-Jack polynomials of 3D Young diagrams which have one layer in $z$-axis direction become the Schur functions of the corresponding 2D Young diagrams, for example, 3-Jack polynomials $\tilde{J}_{\scalebox{0.1}{\includegraphics{3Dfourboxyxy.PNG}}_{h_1,h_2,h_1+h_2}}$ and $\tilde{J}_{\scalebox{0.1}{\includegraphics{3Dfourboxyxy.PNG}}_{h_2,h_1,h_1+h_2}}$ all become Schur function $S_{(2,2)}$. Schur functions of 2D Young diagrams are not dependent on the box growth processes of 2D Young diagrams, while Jack polynomials of 2D Young diagrams and 3-Jack polynomials of 3D Young diagrams are dependent on the box growth processes of 2D/3D Young diagrams. Then we can see that the Littlewood-Richardson rule for $\tilde{J}_{\Dtwoboxy}(\{P_{n,j}\})\times \tilde{J}_{\Dtwoboxx}(\{P_{n,j}\})$ becomes
\[
S_{(2)}S_{(1,1)}=S_{(3,1)}+S_{(2,1,1)}.
\]
We also can check that when $h_1=h, h_2=-h^{-1}$, the Littlewood-Richardson rule for $\tilde{J}_{\Dtwoboxy}(\{P_{n,j}\})\times \tilde{J}_{\Dtwoboxx}(\{P_{n,j}\})$ becomes that for 2D Jack polynomials $\tilde{J}_{(2)} \tilde{J}_{(1,1)}$.

 \section*{Data availability statement}
The data that support the findings of this study are available from the corresponding author upon reasonable request.

\section*{Declaration of interest statement}
The authors declare that we have no known competing financial interests or personal relationships that could have appeared to influence the work reported in this paper.

\section*{Acknowledgements}
This research is supported by the National Natural Science Foundation
of China under Grant No. 12101184 and No. 11871350, and supported by Key Scientific Research Project in Colleges and Universities of Henan Province No. 22B110003.


\begin{thebibliography}{100}
\bibitem{FH}
W. Fulton and J. Harris. {\it Representation theory, A first course}, Springer-Verlag, New York, 1991.
\bibitem{Mac}
I. G. Macdonald. {\it Symmetric functions and Hall polynomials}. Oxford Mathematical
Monographs, Clarendon Press, Oxford, 1979.



\bibitem{weyl}H. Weyl, \emph{The classical groups; their invariants and representations}. Princeton Univ. Press, Princeton, 1946.


\bibitem{MJD}
 T. Miwa, M. Jimbo, and E. Date. \emph{ Solitons: Differential equations, symmetries and infinite dimensional algebras}. Cambridge University Press, Cambridge, 2000.



\bibitem{NVT}
N. Tsilevich.
\emph{Quantum inverse scattering method for the q-boson model and symmetric functions}.
Funct. Anal. Appl. 40, No. 3 (2006) 207-217.



\bibitem{PS} P. Su\l kowski. \emph{Deformed boson-fermion correspondence, Q-bosons, and topological strings on the conifold}. JHEP 10 (2008) 1127-1134.

\bibitem{wang} N. Wang. \emph{Young diagrams in an $N\times M$ box and the KP hierarchy}. Nucl. phys. B  937 (2018) 478-501.

\bibitem{WLZZ}R. Wang, F. Liu, C. H. Zhang, W. Z. Zhao. \emph{Superintegrability for ($\beta$-deformed) partition function hierarchies with $W$-representations}. Eur. Phys. J. C. 82 (2022) 902.

\bibitem{ORV}
A. Okounkov, N. Reshetikhin, C. Vafa. \emph{Quantum Calabi-Yau and classical crystals}. arXiv:hep-th/0309208.

\bibitem{NT}
T. Nakatsu, K. Takasaki. \emph{Integrable structure of melting crystal model with external potentials}. Adv. Stud. Pure  Math. 59 (2010) 201-223.

 \bibitem{Pro1} T. Proch\'{a}zka, {\it Instanton $R$-matrix and $W$-symmetry}, JHEP 12 (2019) 099.

\bibitem{Pro} T. Proch\'{a}zka. \emph{$\mathcal{W}$-symmetry, topological vertex and affine Yangian}. JHEP 10 (2016) 077.
\bibitem{Tsy} A. Tsymbaliuk. \emph{ The affine Yangian of $gl_1$ revisited}. Adv. Math. 304 (2017) 583-645.
\bibitem{FW} O.Foda, M.Wheeler. {\it Hall-Littlewood plane partitions and KP}. 	Int. Math. Res. Not. (2009) 2597-2619.
\bibitem{3DFermionYangian}
N. Wang, K. Wu. \emph{3D Fermion Representation of Affine Yangian}. Nucl. phys. B 974 (2022) 115642.
\bibitem{pro1411} T. Proch\'{a}zka, {\it Exploring $W_{\infty}$ in the quadratic basis}, arXiv: 1411.7697.

\bibitem{3-schur}  N. Wang. \emph{Affine Yangian and 3-Schur functions}. Nucl. phys. B 960 (2020) 115173.
\bibitem{WBCW}
N. Wang, B. Yang, Z. N. Cui, and K. Wu. \emph{Symmetric functions and 3D Fermion representation of $W_{1+\infty}$ algebra}. Adv. Appl. Clifford Algebra 33, 3 (2023).
\bibitem{3-jack}
N. Wang, \emph{3-Jack polynomials and Yang-Baxter equation}, accepted to Reports on Mathematical Physics.
\bibitem{3DbosonYangian}N. Wang, K. Wu. {\it 3D Bosons, 3-Jack polynomials and affine Yangian of ${\mathfrak{gl}}(1)$}, arXiv: 2212.05665.














\end{thebibliography}
\end{document}